\documentclass[twoside,a4paper,11pt]{torus2012}
\usepackage{graphicx}
\usepackage{hyperref}
\usepackage{movie15}
\begin{document}
\pagenumbering{arabic}
\pagestyle{myheadings}
\thispagestyle{empty}
{\flushleft\includegraphics[width=\textwidth,bb=58 650 590 680]{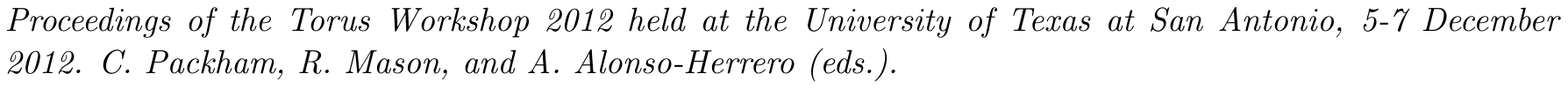}}
\vspace*{0.2cm}
\begin{flushleft}
{\bf {\LARGE
%
The distribution of  AGN covering factors%
}\\
\vspace*{1cm}
%
Andy Lawrence$^{1}$,
Isaac Roseboom$^{1}$,
Jack Mayo$^{1}$,
Martin Elvis$^{2}$, 
Yue Shen$^{2}$, 
Heng Hao$^{2}$, 
and 
Sara Petty$^{3}$
%
}\\
\vspace*{0.5cm}
%
$^{1}$
IfA, Univ.Edinburgh, Royal Observatory, Blackford Hill, Edinburgh EH9 3HJ, UK\\
$^{2}$
Harvard-Smithsonian Center for Astrophysics, 60 Garden St, Cambridge, MA 02138, USA\\
$^{3}$
Physics and Astronomy Dept, Univ.California, Los Angeles, CA 90095-1547, USA
%
\end{flushleft}
%
\markboth{
AGN covering factors
}{ 
%
Lawrence, A et al 
%
}
\thispagestyle{empty}
\vspace*{0.4cm}
\begin{minipage}[l]{0.09\textwidth}
\ 
\end{minipage}
\begin{minipage}[r]{0.9\textwidth}
\vspace{1cm}
\section*{Abstract}{\small
%

We review our knowledge of the most basic properties of the AGN obscuring region - its location, scale, symmetry, and mean covering factor - and discuss new evidence on the distribution of covering factors in a sample of $\sim 9000$ quasars with WISE, UKIDSS and SDSS photometry. The obscuring regions of AGN may be in some ways more complex than we thought - multi-scale, not symmetric, chaotic - and in some ways simpler - with no dependence on luminosity, and a covering factor distribution that may be determined by the simplest of considerations - e.g. random misalignments.
%
\normalsize}
\end{minipage}
%
%
%
\section{Introduction \label{intro}}

Since the 1980s, the standard unified scheme for Active Galactic Nuclei (AGN) has included an optically and  geometrically thick obscuring region, the viewing angle towards which determines whether we see a Type I or Type II AGN \cite{LE82,AM85,KB86,UP95}. Because it must be very thick ($H/R\sim 1$) and is generally assumed to be rotating, the obscuring region has become known as the ``torus'' even though its exact geometry is unclear. Since Krolik and Begelman 1988 \cite{KB88} it has been known that such a structure is hard to maintain, and must be clumpy or filamentary.  Models with a a wedge-like distribution of clumps have had some success in explaining the AGN spectral energy distribution (SED) in the MIR (eg \cite{Nenkova02, AH03, Honig06, Stalevski12}). However such models are purely phenomenological. What have we learned about the actual state of this material? In this conference paper, we look at what the observational evidence tells us about the most basic properties of the obscuring region - its location, scale, symmetry, and covering factor, and especially the {\em distribution} of covering factors.

\section{Location of the obscuring region \label{location}}

Much evidence shows that AGN obscuration is actually a multi-scale, multi-component phenomenon. A good recent summary is in Elvis (2012) \cite{Elvis12}. X-ray absorption is often variable indicating that it occurs on scales of $\leq 10^3 R_S$ although there may also be absorption on ``torus'' scales. The characteristic SED peak in the MIR on the other hand suggests optically thick dust on scales of $\sim 10^5 R_S \sim {\rm pc\;} \times M_H/10^8 M_\odot$ - the traditional ``torus'' component.  MIR interferometry is just begining to probe these scales (\cite{Jaffe04, Tristram12}; see also the contribution in this volume by Tristram). AGN also seem to be routinely accompanied by star-forming discs on a scale of hundreds of parsecs, which can contribute significantly to obscuration (see the contributions by Burtscher and by Imanishi). Finally, the interstellar medium of the host galaxy can be crucial, as originally suggested by Keel (1980) \cite{Keel80}. (See also the contribution by Goulding). Although we loosely refer to ``the torus'', the true situation may be much more complicated than a single neat parsec-scale structure.

\section{Symmetry of the obscuring region \label{symmetry}}

\begin{figure}
\center
\includegraphics[width=0.5\textwidth]{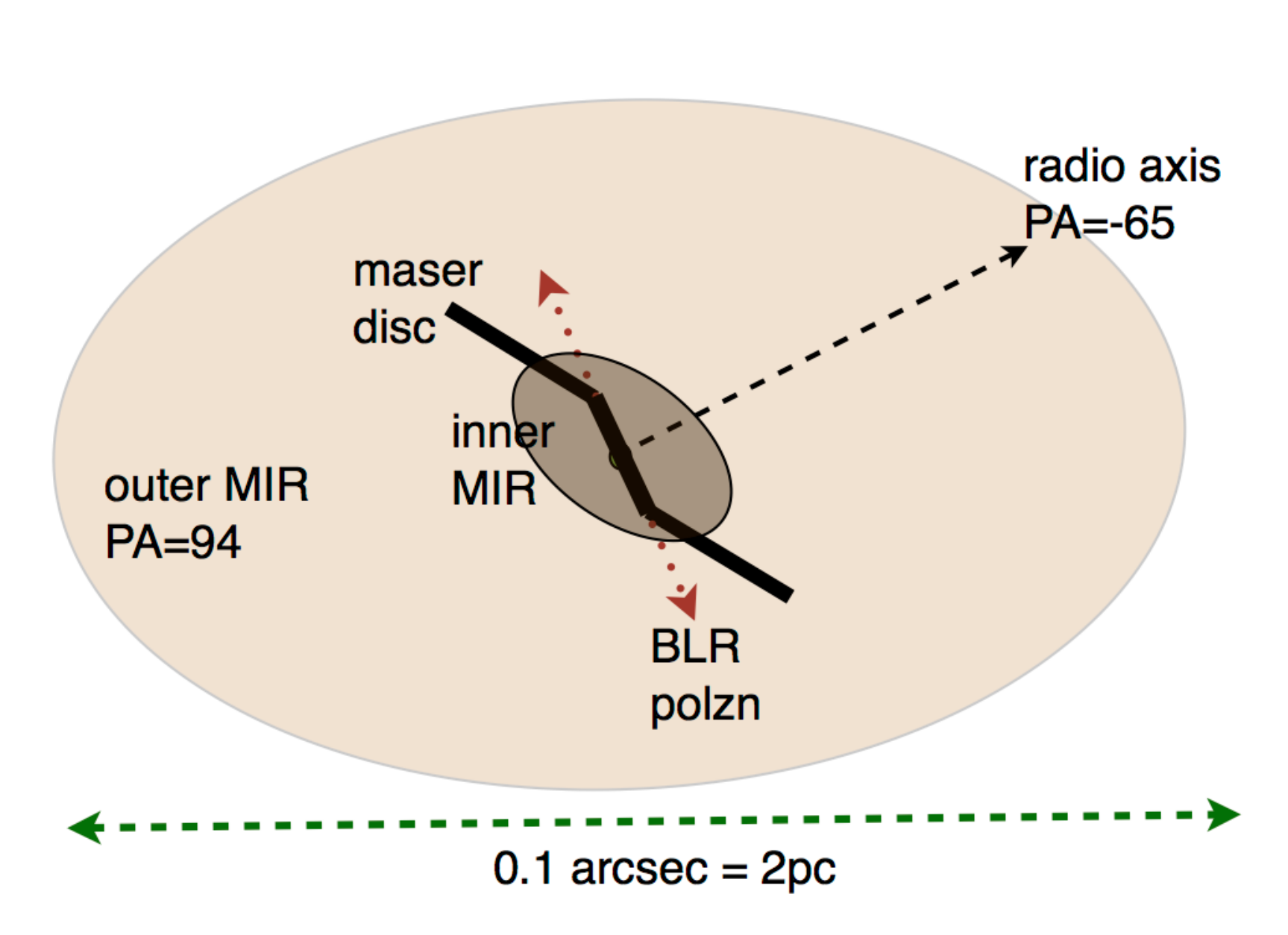} 
\caption{\label{fig:circinus} \it Cartoon indicating structures in the Circinus galaxy on sub-arcsec scales. A large scale radio flow \cite{Elmouttie98} is closely orthogonal to the BLR polarisation \cite{Oliva98}, indicating the true nuclear axis. The inner maser disc \cite{Greenhill03} matches this, but rotates at large radii. The MIR structure \cite{Tristram12} can be modelled with two components, with the outer part at a different PA from the inner part. On a scale of a few arcsec, there is a one-sided OIII cone. The jet axis is inside this cone, but does not bisect it, as shown by \cite{Greenhill03}.
}
\end{figure}

It is generally {\em assumed} that the obscurer is azimuthally symmetric. For a torus supported by rotation, or a dust bearing disc wind, this is what we would expect. But if the material is part of an incoming streamer \cite{Sanchez09} or a warped disc \cite{Greenhill03} it will be not be symmetric. The observational sign of this in projection on the sky would be misalignments between various AGN structures.  Such misalignments in the archetypal AGN NGC 4151, NGC 1068, and Cyg A are discussed by \cite{LE10}. Another example is Cir X-1, a cartoon of which is shown in Fig. \ref{fig:circinus}. Like in NGC1068, the characteristic position angle on the sky seems to change systematically with radial scale, a possible sign of incoming material warping to meet the nuclear axis. It seems that the ``torus'' is not necessarily a neat symmetrical doughnut structure. It is more likely to be chaotic or systematically distorted and asymmetrical.

\section{Average covering factor \label{average}}

There are two ways we can estimate the global covering factor of the obscuring region - by counting AGN types, or by looking at what fraction of the AGN power is re-processed into the IR.  The fraction of obscured AGN has been discussed many times, and recently reviewed by Lawrence and Elvis (2010) \cite{LE10}, who find that the fraction of heavily obscured  objects is $0.58\pm0.03$ and the fraction of lightly obscured objects, where we can still see the BLR, is $0.15\pm0.05$. The latter are probably host-obscured objects, so that the fraction of $\sim 0.6$ relates to the parsec-scale torus-like region.

On the other hand, we know that the IR emission constitutes  around 30\% of the bolometric output of the average AGN \cite{Sanders89, Elvis94}, suggesting that this is the fraction re-processed by the obscuring material. It is interesting that the reprocessed fraction (0.3) agrees roughly with the obscured fraction (0.6), but not exactly. We can only measure the re-processed fraction for Type I AGN; so this is an indication that the typical properties of Type I and II AGN are not the same, a point we return to below.

It is very striking that the obscuring region covers {\em half the sky} as seen from the central engine. As well as being scarily large, and so hard to explain in rotational models, it is a curious number - why not 1\%, or 99\% ? It is a big clue that we should be looking at either disc wind models (eg \cite{Elvis02, Elitzur06}) or warped disc models (eg \cite{Phinney89, LE10}) both of which naturally give rise to covering factors of roughly this order. 

\section{Luminosity dependence of covering factor \label{lum-dep}}

\begin{figure}
\center
\includegraphics[width=0.7\textwidth]{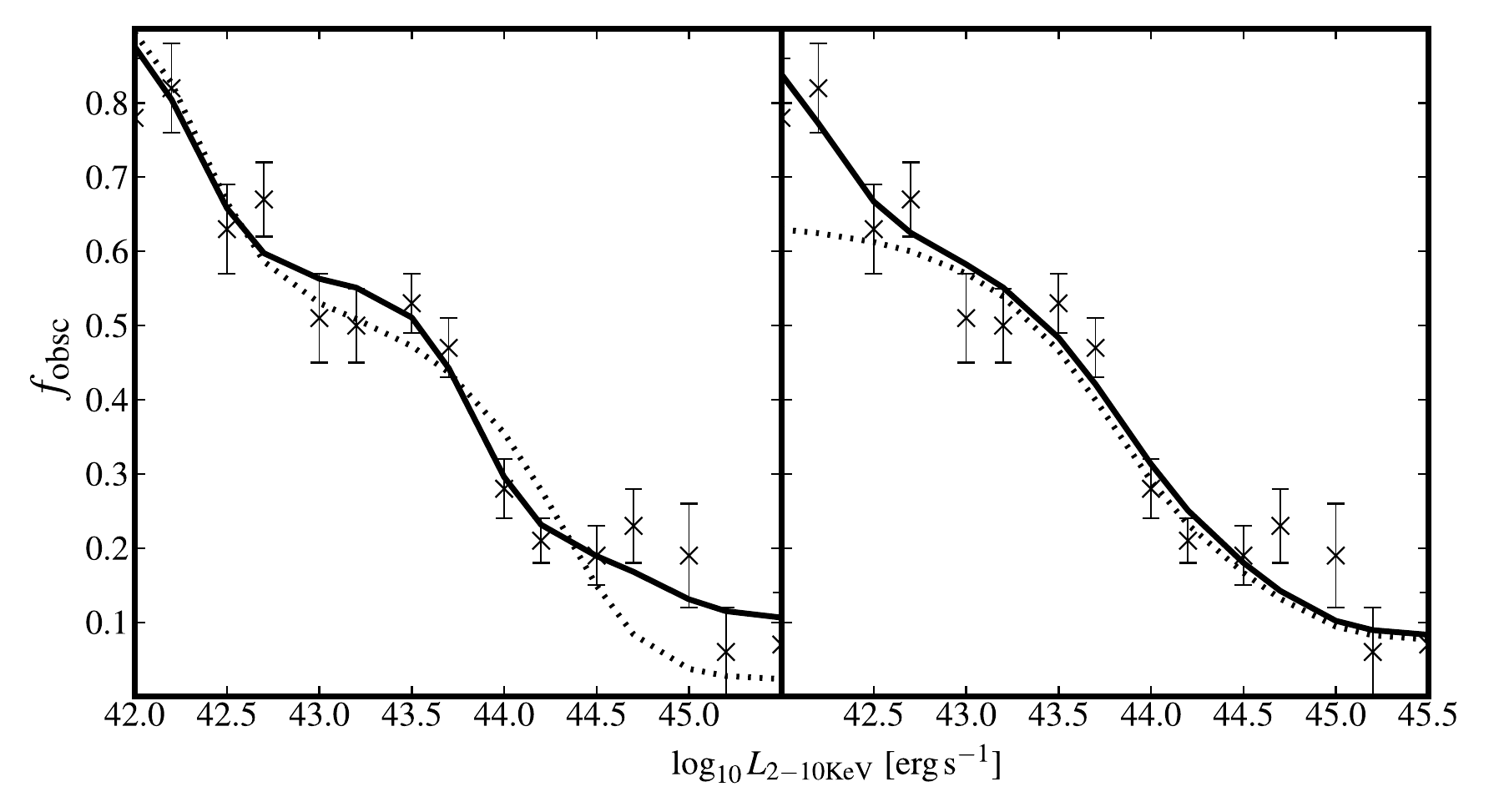} 
\caption{\label{fig:mayo-fc} \it Possible explanation of the luminosity dependence of X-ray obscured fraction, from Mayo and Lawrence in preparation. The data are from \cite{Hasinger08}. The curves represent various models involving populations with varying degrees of partial covering. There is no intrinsic variation of obscured fraction with luminosity; the effect in these models is caused by partial covering biasing the apparent luminosity. 
}
\end{figure}

There have been many discussions of the possible dependence of obscuration on AGN luminosity (eg \cite{LE82, L91, Simpson05, Hasinger08}; see also the contribution by Toba). There are a number of confusing factors and observational problems, such as where to place LINERs and weak-lined radio galaxies, the difficulty of getting good classifications in faint and high redshift samples, and the conflation of optical obscuration and X-ray absorption, which don't always go together. The evidence was reviewed by \cite{LE10}, who came to a puzzling conclusion.  For optical, radio, and IR selected samples, there seems to be no effect, with the fraction of Type II AGN staying the same at roughly 0.6 over several decades of luminosity. However, X-ray samples unambigously show a decline in obscured fraction with increasing luminosity, from 0.8 to 0.1 over 4-5 decades in X-ray luminosity. One possibility is that the X-ray effect is caused by common but unrecognised partial covering by Compton thick material in objects which are mistakenly thought to be absorbed but Compton-thin, which would systematically bias the apparent luminosities of such objects  (Mayo and Lawrence in preparation; see Fig \ref{fig:mayo-fc}).

\section{Probability distribution of covering factor \label{distbn}}

\subsection{Evidence for a range of covering factors \label{range}}

The simplest unified scheme assumes that all AGN are identical, with apparent differences being due to the accident of viewing angle. The simplest variation assumes that all AGN are the same {\em on average}, but that there is a distribution of some properties, for example the covering factor of obscuring material. As first pointed out in \cite{L91}, this will lead to Type I and II AGN having apparently different properties; objects which happen to have larger covering factors, when oriented at random, are more likely to be observed as Type II AGN than objects which have lower covering factors. In general if the probability density distribution of covering factors $C$ is $P(C)$ then Type II objects will have $P(II)= CP(C)$ and Type I objects will have $P(I)=(1-C)P(C)$ .

\begin{figure}
\center
\includegraphics[width=0.5\textwidth]{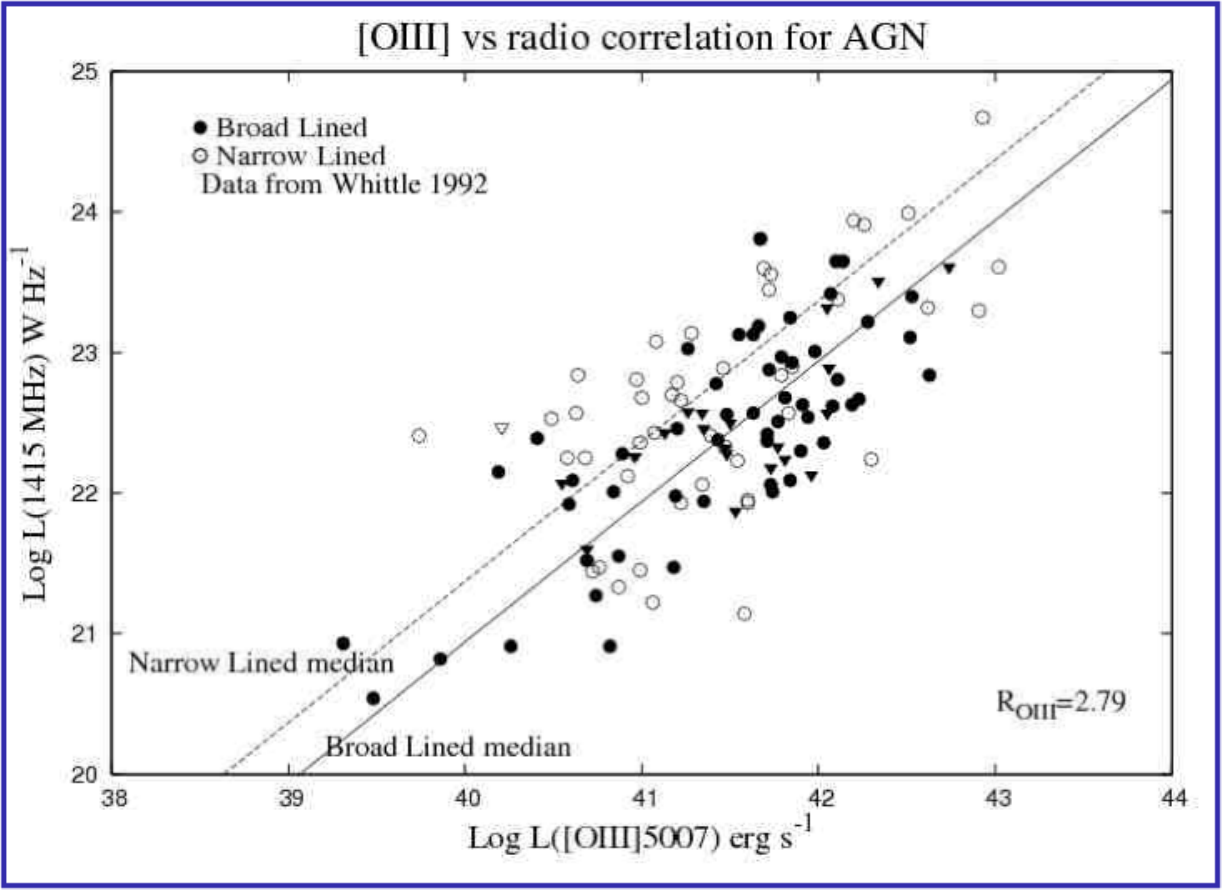} 
\caption{\label{fig:rad-vs-oiii} \it Comparison of radio vs OIII power for Type I and Type II AGN. Taken from \cite{L91}
}
\end{figure}

An example is shown in Fig. \ref{fig:rad-vs-oiii}, which shows that for a given radio power, Type I AGN have stronger OIII emission than Type II AGN. On the assumption that radio power is an indicator of the true UV power, the effect could be produced by a single population because objects with a large covering factor will have a smaller open-cone angle, and so relatively weaker OIII. 
The distribution of covering factors can also potentially explain why the covering factor implied by the IR luminosity of Type I AGN is half as big as that suggested by the Type II/Type I ratio  - objects with smaller covering factors are more likely to observed as Type I AGN. It could also be relevant to the claim that Type II AGN on average have more prominent dust lanes in their circumnuclear regions \cite{Malkan98, Hunt04}. 

\subsection{Predictions for the range of covering factors \label{predictions}}

A range of covering factors can arise fairly naturally in disc wind models, warped disc models, and chaotic accretion models. The only model so far however which has made a concrete prediction for the distribution is the simple misaligned disc model \cite{LE10}. This is only a first step towards a proper physical model, but has the advantage of clear predictions with no free parameters. The hypothesis is that the direction of incoming material is completely random with respect to the nuclear axis, and that the flow warps at some intermediate radius. The first simple variant assumes that during the warp the line of nodes rotates fully. This produces far too many obscured objects. The second simple variant assumes that there is no rotation of the line of nodes; the disc just tilts, with no twist. This predicts $\mu(C_{all})=0.5$, $\mu(C_{I})=0.35$ and $\mu(C_{II})=0.7$. This very crude model roughly gets correct the overall obscured fraction, the typical IR power, and the factor 2 difference in OIII strength.

\subsection{Modeling WISE-UKIDSS-SDSS quasars \label{WUS}}

The WISE survey \cite{Wright11} has given us the oppportunity to construct IR-UV SEDS for large samples of AGN. In \cite{R12} we have used this to study the variation in relative IR to UV strength, and hence covering factor, in luminous quasars. We start with the SDSS-DR7 quasars (\cite{Schneider10, Shen11}; 105,783 objects). We then restrict  to $L_{bol}>10^{46}$ erg s$^{-1}$ and $z<1.5$ (26,927 objects). We then crossmatch with the UKIDSS LAS YJHK survey \cite{L07}, which covers 4000 sq.deg., giving 9,230 objects - all the SDSS quasars are easily seen in UKIDSS. Of these, 9,112 are detected in at least one WISE band, but just 3,831 in all four WISE bands. The latter is our prime sample, but we have carefully modelled selection effects by use of the larger sample.

We then fit our SEDs with three empirical components. (i) The mean optical-UV SED from \cite{Elvis94}, representing the accretion disc;  (ii) hot dust modelled as a single black body; and (iii) a torus component chosen from a library of clumpy torus models \cite{Nenkova08}. We do not intend to extract meaningful physical parameters for the models; they are just an objective way to join the dots and characterise the IR luminosity. We then estimate the covering factor as the ratio of IR luminosity (both components) to the bolometric luminosity.

\begin{figure}
\center
\includegraphics[width=0.6\textwidth]{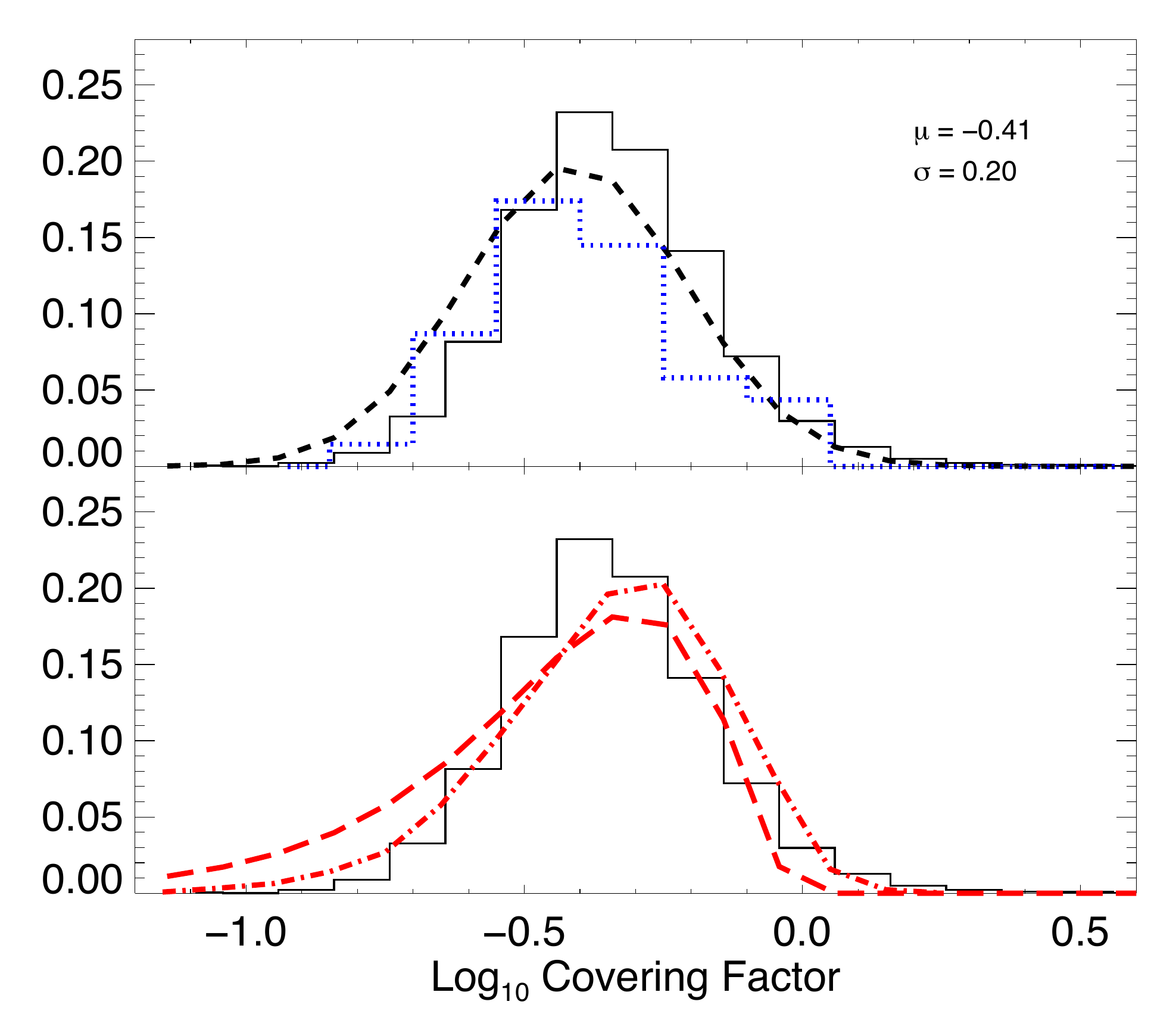} 
\caption{\label{fig:cfac2-all} \it Distribution of luminous quasar covering factors, derived from WISE-UKIDSS-SDSS SEDs, taken from Roseboom et al \cite{R12}. In the upper panel, the black histogram is the raw data, and the dashed curve is the version corrected for incompleteness. The blue histogram is from the sample of \cite{Richards06}. In the lower panel, the red curves are the prediction from the tilted disc model \cite{LE10}; the dot-dashed line includes the selection correction. 
}
\end{figure}

The result is shown in Fig. \ref{fig:cfac2-all}. The observed distribution is approximately a Gaussian in $\log C$ with $\mu=-0.41$ and $\sigma=0.20$, i.e. a mean value $\mu_C=0.39$ and a one sigma range in $C$ of a factor 2.5. This is the distribution in $C$ for Type I AGN; but we can now deduce the distribution for all AGN from which this drawn, which has $\mu_C({\rm all})=0.64$, in fairly good agreement with the observed obscured fraction, which suggests $\mu_C=0.58$. Note that this result is for high luminosity quasars, not low-luminosity AGN; it is in strong disagreement with the Spitzer-based claim of Treister et al (2008) that the covering factor is $\sim 0.04$ at similar bolometric luminosities. Fig. \ref{fig:cfac2-all} also shows the prediction from the simple warped disc model referred to above. It is in good but not perfect agreement. Note that this prediction has {\em no free parameters}. It would be good to see predictions from other possible models, such as disc wind models.

\subsection{Hot dust behaviour \label{hot}}

\begin{figure}
\center
\includegraphics[width=0.95\textwidth]{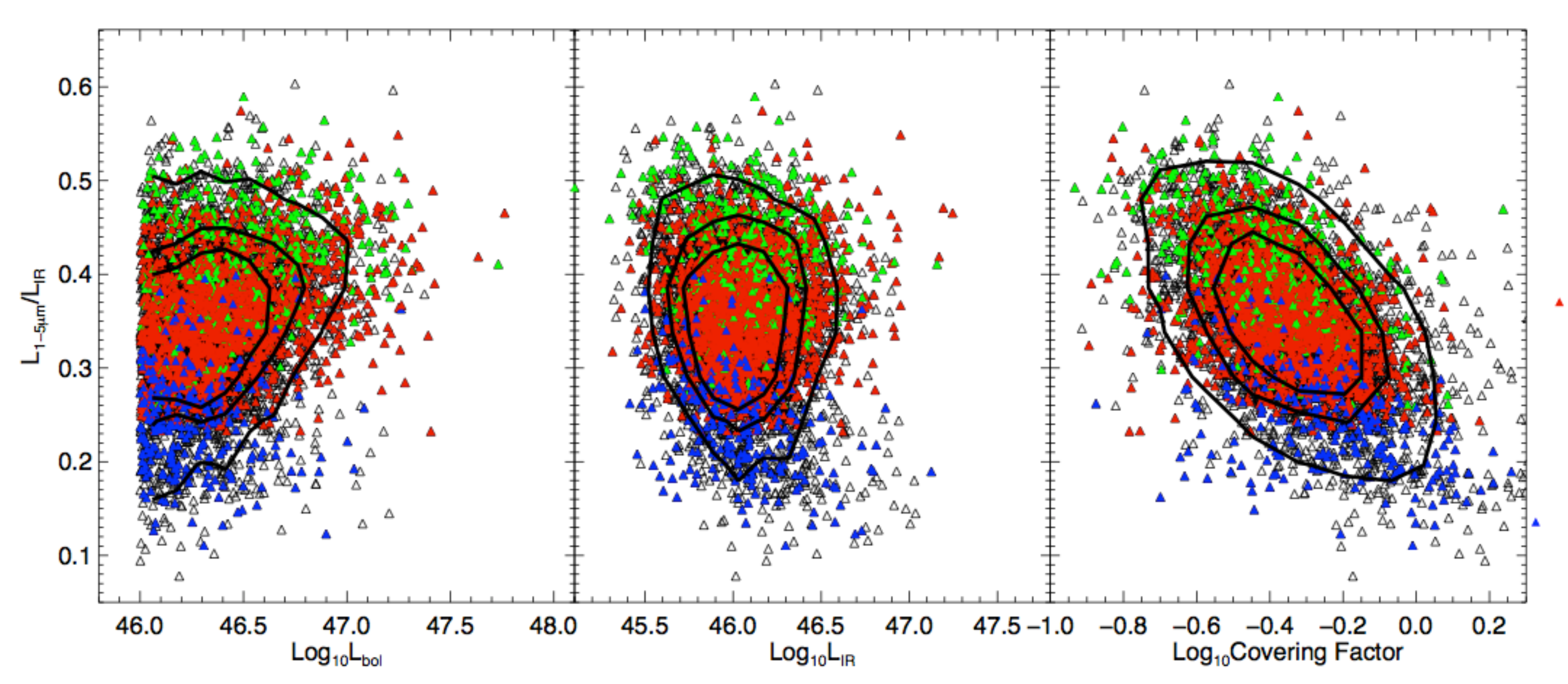} 
\caption{\label{fig:Chot} \it Correlation of the ratio of near-IR and total IR luminosities with various quantities:  bolometric luminosity, IR luminosity, and overall covering factor. Taken from \cite{R12}.
}
\end{figure}

In \cite{R12}, we fitted a hot dust component as well as a cooler torus component. It is clear that as well as the overall covering factor having a range of values, the relative strength of the hot component also varies considerably. Fig. \ref{fig:Chot} shows the behaviour of ratio of the 1-5$\mu$m luminosity to the integrated IR luminosity. Two interesting things stand out. The first is that the contribution of the hot dust is quite large - typically 30\% of the total. This is a significant challenge for clumpy torus models and may require a separate component (see \cite{Mor09, Stalevski12}. The second interesting thing is that the hot dust fraction anti-correlates with overall covering factor - objects with low covering factor have stronger hot dust. We speculate that this may be a geometrical effect. The dust nearest the centre will be the hottest and also the most likely to be optically thick, so that there may bne a kind of hot inner wall. For geometrically thin objects this wall may be preferentially seen face on, whereas for geometrically thick objects it may be seen preferentially edge on, as we are looking down the funnel. 

\section{Conclusions \label{conclusions}}

Our main conclusions are as follows. (i) Obscuration is multi-scale and complex. (ii) The axisymmetry of the obscuring region is unclear. (iii) The mean covering factor is $\sim$0.6. (iv) Evidence on luminosity dependence is conflicting, but it seems most likely that there is no intrinsic dependence of covering factor on luminosity. (v) The distribution of covering factors is roughly log-Gaussian, with two-thirds within a factor of 2.5. (vi) A simple tilted disc model is a good fit to the distribution of covering factors. (viii) Objects with larger covering factor have weaker hot dust. 

%
%
\small  
%
%

%

%
\end{document}